\documentclass[12pt,a4paper]{article}

\usepackage{amssymb,amsfonts,amsmath}

\title{ Classical probability model for Bell inequality}
\author{Andrei Khrennikov\\
International Center for Mathematical Modeling\\ 
in Physics, Engineering, Economics, and Cognitive Science\\
Linnaeus University, Sweden}

\begin{document}
\maketitle

\abstract{We show that by taking into account randomness of realization of experimental contexts 
it is possible to construct common Kolmogorov space for data collected for these contexts, 
although they can be incompatible. We call such a construction "Kolmogorovization" of contextuality. 
This construction of common probability space is applied to Bell's inequality.
It is well known that its violation is a consequence of collecting statistical data in a few incompatible experiments. In experiments performed 
in quantum optics contexts are determined by selections of pairs of angles $(\theta_i, \theta^\prime_j)$ fixing orientations
of polarization beam splitters. Opposite to the common opinion, we show that statistical data corresponding to measurements of polarizations 
of photons in the singlet state, e.g., in the form of correlations, can be described in the classical probabilistic framework. The crucial
point is that in constructing the common probability space one has to take into account not only randomness of the source (as Bell did),
but also randomness of context-realizations (in particular, realizations of pairs of angles $(\theta_i, \theta^\prime_j)$). One may (but need not)
say that randomness of ``free will'' has to be accounted.}

\section{Introduction}

The aim of this note is to clarify some details  of author's  construction of classical Kolmogorov probability space 
for the EPR-Bohm correlations which first time was  given in the work \cite{Avis}. Unfortunately, the paper \cite{Avis}
suffered  of a few presentational problems. The main of them is that there were presented two totally different classical Kolmogorov models
for violation of Bell's inequality, one due to D. Avis and another due to the author (and the latter is reconsidered in the present note).
The reader might mix these models into one which induces misunderstandings.

\medskip

The role of Bell's inequality \cite{Bell} for quantum foundations and nowadays for quantum technologies was discussed in hundreds of papers,
see, e.g., \cite{KHR_CONT} for the extended bibliography. Recently crucial experimental tests closing the fair 
sampling loophole were performed \cite{Zeilinger}--\cite{Kwiat2}. Leading experimenters promise that the final (clean and loophole-free) 
test will be performed in a few years, may 
be even next year. Thus it seems that the ``Bell inequality epic'' \cite{Bell}--\cite{Raedt1}  
is near the end. Nevertheless, the probabilistic structure of the Bell argument
has not yet been completely clarified. And the present paper is a step towards such clarification. In physics violation of Bell's inequality
is typically coupled to notions of {\it realism, locality, and free will} \cite{Bell}, \cite{Zeilinger}--\cite{Fine}. In mathematically oriented literature violation of Bell's inequality
is considered as exhibition of {\it nonclassicality of quantum probability} - impossibility to use the 
Kolmogorov model of probability theory \cite{K}. There can be found numerous publications on this topic, see, e.g., 
\cite{KHR_CONT}, \cite{Larsson}--\cite{Marian}, \cite{j1}--\cite{j8}, \cite{L2}--\cite{Raedt1}.
In particular, in a series of works, e.g., \cite{KHR_CONT}, \cite{j1}, \cite{KH_JA},
\cite{ENTROPY},  the author of this paper presented the viewpoint that violation of Bell's inequality is a consequence of combining in one inequality of 
statistical data collected for a few incompatible experimental contexts. (In experiments performed 
in quantum optics contexts are determined by selections of pairs of angles $(\theta_i, \theta^\prime_j)$ fixing orientations
of polarization beam splitters.) Data for each context can be described by the classical probability
space, Kolmogorov space, but there is no common space for this data \cite{KHR_CONT}. This viewpoint was strongly supported by the famous theorem of 
A. Fine \cite{Fine7} coupling Bell's inequality with existence of the joint probability distribution for all measured random variables. 

What would N. Bohr say about such a viewpoint? He would definitely, cf. \cite{BR}, 
say that there is no reason for violation of classicality even if data were collected for incompatible contexts. For him, measurement 
apparatus is a classical device and even if one collects outputs of a few classical devices,  there is no reason to expect a violation of laws
of classical probability. In this note we confirm such a position. We show that by taking into account the randomness of 
realization of experimental contexts it is possible to construct the common Kolmogorov space for data collected for these contexts, 
although they can be incompatible. We call such a construction "Kolmogorovization" of contextuality. 
This construction of common probability space is applied to Bell's inequality.

Opposite to the common opinion, we show that {\it statistical data corresponding to measurements of polarizations 
of photons in the singlet state, e.g., in the form of correlations, can be described in the classical probabilistic framework.} The crucial
point is that in constructing the common probability space one has to take into account not only randomness of the source (as Bell did),
but also randomness of context-realizations; in particular, realizations of the pairs of the angles $(\theta_i, \theta^\prime_j)$ determining orientations 
of polarization beam splitters (PBSs).  Roughly speaking randomness of (pseudo)-random generators controlling selections of 
orientations of PBSs has be added to randomness of the source. Although everybody recognizes that randomness of these generators plays 
an important role in the Bell framework, its presence is ignored in theoretical constructions representing the EPR-Bohm-Bell 
correlations. These generators are viewed as some ``technicalities'' which are interesting for experimenters, but not theoreticians. Recently the author started 
to pay serious attention to various experimental technicalities, to embed them into theoretical models, to lift them to the level of fundamental 
studies, e.g., \cite{KH_JA}. The present paper can be consider as a further step in this direction.

In the Kolmogorov model constructed in this paper {\it quantum probabilities and correlations are represented as conditional probabilities 
and correlations} (under condition 
of fixed experimental settings). Such conditional probabilities and correlations can violate Bell's inequality, 
although unconditional ones have to satisfy it.  We remark that, as was shown in numerous experiments, theoretical quantum probabilities and correlations
 coincide (of course, approximately) with probabilities and  correlations found by experimenters for the fixed experimental settings, 
e.g., the angles $(\theta_i, \theta^\prime_j)$
in test of Bell type inequalities. Therefore our classical probabilistic model represents experimental probabilities and corellations as well.  

Our model is {\it local.} However, it is {\it not objective (realistic)} in the sense of  
Einstein, Podolsky, Rosen and Bell: the values of observables are not determined by the state of a system emitted by 
a source, by, so to say, a hidden variable $\lambda.$  However, 
from Bohr's viewpoint one cannot expect such type of objectivity, since it is defined solely in terms of the states of systems 
emitted by a source. And our model is {\it objective in a more general sense} (which might satisfy Bohr): the hidden variable $\lambda$
and the random parameters in random generators selecting the experimental settings  determine the values of observables 
(e.g., spin or polarization projections).     We remark that even in classical physics randomness of measurement
devices is routinely taken into account and if statistical data is collected by using a few different devices is has to be weighted. Kolmogorovization
of Bell's argument is reduced to such a standard procedure. One may (but need not)
say that randomness of ``free will'' has to be accounted, but we postpone the discussion on this topic, see section \ref{free}.

The main problem handled in this note is following. 
There are a few experimental contexts and  statistical data  (of any origin) collected  
for these contexts. {\it Is it possible to construct common probability space representing 
these data?} We show that the answer is always positive! 

This result changes the viewpoint on the role of quantum probabilistic calculus. As was mentioned, the common viewpoint is that 
we use quantum probabilistic rules, because classical probabilistic rules are violated, see, e.g., 
Feynman \cite{Feynman} and the author \cite{KHR_CONT}. (Besides violation of Bell's inequality, there is typically stressed that 
the formula of total probability is violated by statistical data collected in the two slit and other interference experiments.) In this 
note we show that such an ``impossibility viewpoint'' has to be changed. In the light of 
Kolmogorovization it is clear that the calculus of quantum probabilities is just a calculus of conditional probabilities defined in 
the classical probabilistic framework. It is not surprising that conditional probabilities, although generated in the common probability space, 
but with respect to different conditions, follow rules which are different from the rules which were derived for unconditional probabilities,
such as, e.g., Bell type inequalities.

\section{CHSH-inequality}

We recall the rigorous mathematical formulation of the CHSH
inequality:

{\bf Theorem 1.} {\it Let $A^{(i)}(\omega)$ and $B^{(i)}(\omega), i=
1,2,$ be random variables taking values in $[-1, 1]$ and defined
on  a single probability space ${\cal P}.$ Then the following
inequality holds:}
\begin{equation}
\label{CHSH} |<A^{(1)},B^{(1)}> + <A^{(1)},B^{(2)}> + <A^{(2)},
B^{(1)}> - <A^{(2)},B^{(2)}>| \leq 2.
\end{equation}

Correlation is defined as it is in classical
probability theory:
$$
<A^{(i)},B^{(j)}>= \int_\Omega A^{(i)}(\omega) B^{(j)}(\omega) d
{\bf P} (\omega).
$$

Experimental tests of the CHSH-inequality are based on the following methodology. 
One should put statistical data collected
for four pairs of PBSs settings:
$$
\theta_{11}=(\theta_1, \theta_1^\prime), \theta_{12}=(\theta_1,
\theta_2^\prime), \theta_{21}=(\theta_2, \theta_1^\prime),
\theta_{22}=(\theta_2, \theta_2^\prime),
$$
into it. Here $\theta= \theta_1, \theta_2$ and $\theta^\prime=
\theta_1^\prime, \theta_2^\prime$ are selections of angles  for
orientations of respective PBSs.

  Following Bell, the selection of the angle $\theta_i$ determines the
random variable
$$
a_{\theta_i}(\omega)= \pm 1.
$$
There are two detectors coupled to the PBS with the
$\theta$-orientation: "up-spin" (or "up-polarization") detector
and "down-spin" (or "down-polarization") detector.  
A click of the up-detector assigns to the random variable $a_{\theta}(\omega)$
the value +1  and a click of the down-detector assigns to it the
value -1. 
In the same way selection of the angle $\theta^\prime$ determines
$$
b_{\theta_i^\prime}(\omega)= \pm 1.
$$
If one assumes that these observables can be represented as random variables on common Kolmogorov probability space, then
their correlations have to satisfy the CHSH-inequality (\ref{CHSH}). However, the correlations calculated with the aid of the 
quantum formalism as well as experimental correlations violate this inequality. Therefore the above assumption about Kolmogorovness 
of data has to be rejected, see A. Fine \cite{Fine7}, see A. Khrennikov \cite{j1},  \cite{KHR_CONT} for discussions.  
In this paper we do not question this conclusion. Our point is that from the classical probabilistic viewpoint there is 
no reason to assume that these  correlations can be reproduced in common probability space. However, they can be embedded in 
 ``large Kolmogorov space'' as conditional correlations.

\subsection{Random experiment taking into account random choice of setting}

\begin{itemize}
\item[a).] There is a source of entangled photons. 

\item[b).] There are two pairs of PBSs with the corresponding pairs of detectors. PBSs in pairs are
oriented with angles numbered $i= 1,2$ and $j=1,2.$ The pairs of PBSs (with their detectors) 
are located in spatially separated labs, say Lab1 (PBSs $i=1,2)$  and Lab2  (PBSs $j=1,2).$ 
The source is connected (e.g., by optical fibers) to the Labs.

\item[c).] In each lab there is a distribution device;
at each instance of time\footnote{Timing can be experimentally realized with the aid of the cell method used in \cite{Kwiat1}.}:
 $t=0, \tau, 2\tau, \ldots,$ it opens the gate 
to only one of the two channels going to  the
corresponding PBSs. For simplicity, we suppose that in each Lab channels are opened with equal
probabilities
$$
{\bf P}(i)= {\bf P}(j)=1/2.
$$
\end{itemize}
  
Now in each of labs we define two observables corresponding to PBSs (with their detectors):

\begin{itemize}
\item[A1.]  $A^{(i)}(\omega)= \pm 1, i=1,2$ if in Lab1 the $i$-th channel is open and the corresponding (up or
down)  detector coupled to $i$th PBS fires;

\item[A0.] $A^{(i)}(\omega)= 0$ if in Lab1  the $i$-th channel is  blocked. 

\item[B1.]  $B^{(j)}(\omega)= \pm 1, j=1,2$ if in Lab2 the $j$-th channel is open and the corresponding (up or
down)  detector coupled to $j$th PBS fires;

\item[B0.] $B^{(j)}(\omega)= 0$ if in Lab2  the $j$-th channel is  blocked. 
\end{itemize}

We remark that in the present experiments testing the CHSH-inequality 
experimenters do not use four PBSs, but only two PBSs (one at each side);
not two pairs of detectors at each side, but just one pair. The random choice 
of orientations $\theta=\theta_1, \theta_2$ (on one side) and $\theta^\prime=\theta^\prime_1, \theta^\prime_2$
(on another side) is modeled with the aid of additional devices preceding the corresponding PBSs. 
(Pseudo)-random generators specify parameters in these devices corresponding to selection of two 
different orientations for each of two PBSs. The main reason for the use of this scheme is that it is
simpler in realization and it is essentially cheaper. The latter is very important, since photo-detectors 
with approximately 100\% efficiency are extremely expensive. There are no doubts that everybody would agree 
that the two experimental schemes under discussion represent the same physical situation. 
However, the scheme with four PBSs is more natural from the probabilistic viewpoint. At each side there 
are two pairs of detectors determining two random variables. Roughly speaking there are two observers 
at each side, Alice1, Alice2 and Bob1, Bob2,  each of them monitors her/his pair of detectors and,  
for each emitted system, she/he has to assign some value, if non of the detectors clicks she/he assigns 0.       
(We consider the ideal situation without losses).

\subsection{Kolmogorovization of incompatible statistical data}

We now construct a proper Kolmogorov probability space for the
EPR-Bohm-Bell experiment. 

This is a general construction for
combining of probabilities produced by a few incompatible
experiments. 

For the fixed pair of orientations $(\theta_i,\theta_j^\prime),$ there are given probabilities $p_{ij}(\epsilon,
\epsilon^\prime), \epsilon, \epsilon^\prime= \pm 1, $ to get the values 
$(\epsilon, \epsilon^\prime).$ These are either experimental probabilities (frequencies) or probabilities
produced by the mathematical formalism of QM. For the singlet state and measurements of polarization, we have:      
\begin{equation}
\label{QM} 
p_{ij}(\epsilon, \epsilon)=\frac{1}{2} \cos^2
\frac{\theta_i-\theta_j^\prime}{2}, p_{ij} (\epsilon,
-\epsilon)=\frac{1}{2} \sin^2 \frac{\theta_i-\theta_j^\prime}{2}.
\end{equation}

However, this special form of probabilities is not important for
us. Our construction of unifying Kolmogorov probability space
works well for any collection of numbers $p_{ij}(\epsilon,
\epsilon^\prime)$ such that for any pair $(i,j):$
$$
0\leq p_{ij}(\epsilon,
\epsilon^\prime)\leq 1, \; 
\sum_{\epsilon, \epsilon^\prime} p_{ij} (\epsilon,
\epsilon^\prime)=1.
$$ 
Let us now consider the  set of points $\Omega:$ 
$$
\omega= (\epsilon_1, 0, \epsilon_1^\prime, 0), (\epsilon_1, 0, 0,
\epsilon_2^\prime), (0, \epsilon_2, \epsilon_1^\prime, 0), (0,
\epsilon_2, 0, \epsilon_2^\prime),
$$
where $\epsilon =\pm 1, \epsilon^\prime= \pm 1.$
These points correspond to the following events:
e.g., $(\epsilon_1, 0, \epsilon_1^\prime, 0)$
means: at the left hand side PBS N 1 is coupled and PBS N 2 is uncoupled and the same is 
at the right hand side, the result of measurement at the left hand side 
after PBS N 1 is given by $\epsilon_1$ and at the right hand side  by $\epsilon_1^\prime.$

We define the following probability measure
on~$\Omega:$
$${\bf P}(\epsilon_1, 0, \epsilon_1^\prime, 0) = \frac{1}{4} p_{11}(\epsilon_1, \epsilon_1^\prime),
{\bf P}(\epsilon_1, 0, 0, \epsilon_2^\prime) = \frac{1}{4}
p_{12}(\epsilon_1, \epsilon_2^\prime)
$$
$$
{\bf P}(0, \epsilon_2, \epsilon_1^\prime, 0) = \frac{1}{4}
p_{21}(\epsilon_2, \epsilon_1^\prime), {\bf P}(0, \epsilon_2, 0,
\epsilon_2^\prime) = \frac{1}{4} p_{22}(\epsilon_2,
\epsilon_2^\prime).
$$

We now define random variables $A^{(i)}(\omega), B^{(j)}
(\omega):$
$$
A^{(1)}(\epsilon_1, 0, \epsilon_1^\prime, 0)= A^{(1)}(\epsilon_1,
0, 0, \epsilon_2^\prime)= \epsilon_1, A^{(2)}(0, \epsilon_2,
\epsilon_1^\prime, 0)= A^{(2)}(0, \epsilon_2, 0,
\epsilon_2^\prime)= \epsilon_2;
$$
$$
B^{(1)}(\epsilon_1, 0, \epsilon_1^\prime, 0)= B^{(1)}(0,
\epsilon_2, \epsilon_1^\prime, 0)= \epsilon_1^\prime,
B^{(2)}(\epsilon_1, 0, 0, \epsilon_2^\prime)= B^{(2)}(0,
\epsilon_2, 0, \epsilon_2^\prime)=\epsilon_2^\prime.
$$
and we put these variables equal to zero in other points; for example,
$A^{(1)}(0, \epsilon_2, \epsilon_1^\prime, 0)=  A^{(1)}(0,
\epsilon_2, 0, \epsilon_2^\prime)=0.$ Thus if the channel going to PBS N1
at Alice1/Alice2 side is closed, then, since non of detectors following it fires, Alice1 
assigns the value 0 to her observable $A^{(1)}.$  

These random variables are local in the sense that their values do not depend 
on experimental context and the results of measurements on the opposite side;
for example,  $A^{(1)}(\omega)$ depends only on the first coordinate of $\omega= 
(\omega_1, \omega_2, \omega_3, \omega_4).$

We find two dimensional probability distributions; nonzero are given by  
$${\bf P}(\omega \in \Omega:
A^{(1)} (\omega)= \epsilon_1, B^{(1)}(\omega)= \epsilon^\prime_1)=
{\bf P}(\epsilon_1, 0, \epsilon_1^\prime, 0)= \frac{1}{4} p_{11}
(\epsilon_1, \epsilon_1^\prime), \ldots,
$$
$$
{\bf P} (\omega \in \Omega: A^{(2)} (\omega)= \epsilon_2,
B^{(2)}(\omega)= \epsilon_2^\prime) = \frac{1}{4} p_{22}
(\epsilon_2, \epsilon_2^\prime).
$$
Then, e.g., 
$$
{\bf P}(\omega \in \Omega:
A^{(1)} (\omega)= \epsilon_1, A^{(2)}(\omega)= \epsilon_2) = {\bf P}(\emptyset)= 0.
$$
We also consider the random variables which monitor selection of channels:
$\eta_a=i, i=1,2$  if the channel to the $i$th PBSs on the Alice1/Alice2 side
is open and $\eta_b= j, j=1,2$  if the channel to the $j$th PBSs on the Bob1/Bob2 side
is open. Thus 
$$
\eta_a(\epsilon_1, 0, 0,\epsilon_2^\prime)= 1, \eta_a(\epsilon_1, 0, \epsilon_1^\prime, 0)=
1, \; \eta_a(0, \epsilon_2, 0, \epsilon_2^\prime)= 2, \eta_a(0, \epsilon_2,
\epsilon_1^\prime, 0)= 2, 
$$
$$
\eta_b(\epsilon_1, 0, \epsilon_1^\prime, 0)=1,  \eta_b(0, \epsilon_2,
\epsilon_1^\prime, 0)= =1, \;  \eta_b(\epsilon_1, 0, 0,\epsilon_2^\prime)= 2,
 \eta(0, \epsilon_2, 0, \epsilon_2^\prime)= 2.  
$$
Here 
$$
{\bf P} (\omega \in \Omega: \eta_a(\omega)=1) =\sum_{\epsilon_1, \epsilon_2^\prime}{\bf P}
(\epsilon_1, 0, 0,\epsilon_2^\prime) +  \sum_{\epsilon_1, \epsilon_1^\prime} {\bf P}
(\epsilon_1, 0, \epsilon_1^\prime, 0)
$$
$$
= 1/4 \Big[\sum_{\epsilon_1, \epsilon_2^\prime} p_{12} (\epsilon_1, \epsilon_2^\prime) +
 \sum_{\epsilon_1, \epsilon_1^\prime} p_{11}(\epsilon_1, \epsilon_1^\prime)\Big] =1/2.
$$
In the same way ${\bf P} (\omega \in \Omega: \eta_a(\omega)=2)= 1/2;
{\bf P} (\omega \in \Omega: \eta_b(\omega)=1) = {\bf P} (\omega \in \Omega: \eta_b(\omega)=2)= 1/2.
$ 
We remark that the random variables $\eta_a$ and $\eta_b$ are independent, e.g., 
$$
{\bf P} (\omega \in \Omega: \eta_a(\omega)=1, \eta_b(\omega)=1)= {\bf P}(\epsilon_1, 0, \epsilon_1^\prime, 0)
= 1/4\sum_{\epsilon_1, \epsilon_1^\prime} p_{11} (\epsilon_1, \epsilon_1^\prime) =1/4
$$
$$
=  {\bf P} (\omega \in \Omega: \eta_a(\omega)=1) {\bf P} (\omega \in \Omega: \eta_b(\omega)=1) = 1/4.
$$
In the same way 
$$
{\bf P} (\omega \in \Omega: \eta_a(\omega)=1, \eta_b(\omega)=2)= {\bf P}(\epsilon_1, 0, 0,\epsilon_2^\prime))
= 1/4\sum_{\epsilon_1, \epsilon_2^\prime} p_{12} (\epsilon_1, \epsilon_1^\prime) =1/4
$$
$$
=  {\bf P} (\omega \in \Omega: \eta_a(\omega)=1) {\bf P} (\omega \in \Omega: \eta_b(\omega)=2) = 1/4
$$
and so on.
These random variables are local in the sense that their values do not depend 
on experimental context and the results of measurements on the opposite side;
for example,  $\eta_a(\omega)$ depends only on the first coordinate of $\omega= 
(\omega_1, \omega_2, \omega_3, \omega_4).$
 
We now find conditional probabilities for the results of joint measurements of observables
$A^{(i)}(\omega)$ and $B^{(j)} (\omega), i, j=1,2,$ conditioned on opening of channels going 
to corresponding PBSs. The definition of the classical (Kolmogorov, 1933) conditional probability is based on Bayes 
formula, e.g., 
$$
{\bf P}(\omega \in \Omega:
A^{(1)} (\omega)= \epsilon_1, B^{(1)}(\omega)= \epsilon_2^\prime \vert \eta_a=1, \eta_b=1) 
$$
$$= 
\frac{{\bf P}(A^{(1)} (\omega)= \epsilon_1, B^{(1)}(\omega)= 
\epsilon_2^\prime,  \eta_a (\omega)=1, \eta_b(\omega)=1)}{{\bf P}(\omega \in \Omega: \eta_a (\omega)=1, \eta_b(\omega)=1)}
$$
$$
= \frac{{\bf P}(\epsilon_1, 0, \epsilon_1^\prime, 0)}{1/4}= p_{11}(\epsilon_1, \epsilon_1^\prime).
$$
In general, we obtain:
$$
{\bf P}(\omega \in \Omega:
A^{(i)} (\omega)= \epsilon_i, B^{(j)}(\omega)= \epsilon_j^\prime \vert \eta_a=i, \eta_b=j) = p_{ij} (\epsilon_i, \epsilon_j^\prime). 
$$
In particular, if initially the weights $p_{ij} (\epsilon_i, \epsilon_j^\prime)$ were calculated with the aid of the rules of quantum mechanics,
for the singlet state and polarization observables, i.e., they have the meaning of quantum probabilities, 
then the classical  conditional probabilities coincide with these quantum probabilities. In this way all quantum probabilities 
can be represented classically. However, they have to be treated as conditional probabilities. In principle, such a viewpoint 
on quantum probabilities is well established, see, e.g., \cite{DEM4},\cite{DEM}, \cite{KHR_CONT}. The novel contribution is that concrete
classical probabilistic construction for embedding of quantum probabilities into classical Kolmogorov model was presented.  

To complete the picture of conditioning, we also present probabilities for observations for  closed channels, e.g.,
$$
{\bf P}(\omega \in \Omega: A^{(1)} (\omega)= \epsilon_1, B^{(1)}(\omega)= \epsilon_2^\prime \vert \eta_a=2, \eta_b=1)=
$$
$$= 
\frac{{\bf P}(A^{(1)} (\omega)= \epsilon_1, B^{(1)}(\omega)= 
\epsilon_2^\prime,  \eta_a (\omega)=2, \eta_b(\omega)=1}{{\bf P}(\omega \in \Omega: \eta_a (\omega)=2, \eta_b(\omega)=1)}
$$
$$
= \frac{{\bf P}(\emptyset)}{1/4}= 0.
$$

\section{Classical conditional probability viewpoint on violation of the CHSH-inequality} 
 
Since the CHSH-inequality is an inequality for correlations, we find them. We are interested in two types of correlations,
so to say, absolute, $E(A^{(i)} B^{(j)}),$ and conditional, $E(A^{(i)} B^{(j)}\vert \eta_a=i, \eta_b=j).$ We have:
$$
E(A^{(i)} B^{(j)}) = \sum_{\epsilon_i, \epsilon_i^\prime} \epsilon_i \epsilon_j^\prime {\bf P} (A^{(i)}=\epsilon_i, 
B^{(j)}= \epsilon_j)  = (1/4)\sum_{\epsilon_i, \epsilon_i^\prime} \epsilon_i \epsilon_j^\prime p_{ij}(\epsilon_i, \epsilon_j^\prime)
$$ 
and 
$$
E(A^{(i)} B^{(j)}\vert \eta_a=i, \eta_b=j)= \sum_{\epsilon_i, \epsilon_i^\prime} \epsilon_i \epsilon_j^\prime {\bf P} (A^{(i)}=\epsilon_i, 
B^{(j)}= \epsilon_j \vert \eta_a=i, \eta_b=j) 
$$
$$
=  
\sum_{\epsilon_i, \epsilon_i^\prime} \epsilon_i \epsilon_j^\prime p_{ij}(\epsilon_i, \epsilon_j^\prime).
$$
Suppose that the weights $p_{ij}$ are selected as quantum probabilities for the singlet state, denote the corresponding 
correlations $C_{ij},$
then 
$$
E(A^{(i)} B^{(j)}\vert \eta_a=i, \eta_b=j) = C_{ij}
$$
and  
$$
E(A^{(i)} B^{(j)}) = C_{ij}/4.
$$
The classical correlations satisfy to conditions of Theorem 1 and, hence, to the CHSH-inequality, i.e.,
the conditional correlations satisfy to the inequality:
\begin{equation}
\label{GHT}
\vert C_{11} + C_{12} + C_{21} - C_{22} \vert \leq 8.  
\end{equation}
Since the conditional correlations coincide with the quantum correlations, we obtain an inequality 
which has to be satisfied by the quantum correlations. However, in the right-hand side there is not 2
as in the CHSH, but 8 and, hence, no problem arises. 

Roughly speaking the bound 8 is too high to make any meaningful constraint on correlations. 
Thus, in this situation the CHSH-inequality gives too rough estimate, since
it is clear that even conditional correlations of random variables with values
$\{-1, 0, +1\}$ are bounded by 1. Thus the worst straightforward estimate has to be
\begin{equation}
\label{GHT1}
\vert C_{11} + C_{12} + C_{21} - C_{22} \vert \leq 4.  
\end{equation}
We now go another way around and from the inequality  (\ref{GHT1}) for conditional correlations we obtain 
an inequality for unconditional (``absolute'') correlations:
\begin{equation}
\label{GHT1}
\vert E(A^{(1)} B^{(1)}) + E(A^{(1)} B^{(2)}) + E(A^{(2)} B^{(1)}) - E(A^{(2)} B^{(2)}) \vert \leq 1. 
\end{equation}

\section{Conclusion} 

We showed that one can easily violate the CHSH-inequality in the classical probabilistic framework (the Kolmogorov model \cite{K}).

We emphasize that a violation of the CHSH-inequality for quantum correlations represented as classical conditional correlations
is not surprising.  In our approach the main problem is to explain Tsirelson bound. 

We repeat our interpretation of the obtained result. Our model is local. It seems that ``nonlocality'' is 
an artificial issue without direct relation to violation of Bell's type inequalities. However, our model is not realistic
in the sense of naive realism of Einstein, Podolsky, Rosen, and Bell, i.e., the measured values cannot be predetermined on the 
basis of the state of a system. Nevertheless, a kind of contextual realism (of Bohr's type) is preserved: by taking into account
not only the state of a system, but also randomness of (pseudo)-random generators selecting orientations of PBSs, it is possible to violate
the CHSH-inequality.

\section{Appendix: Taking into account free will?}
\label{free}

Recently so-called free will problem became an important topic in  quantum foundational discussions. In particular, the recent spike 
of activity was generated by `t Hooft's position that quantum mechanics can be consistently explained by using the totally deterministic 
picture of nature \cite{Hooft}. Of course, anybody having elementary education in cognitive science and philosophy would consider 
the ``free will'' discussions in quantum foundations as very  primitive, cf. with \cite{Free}. They have practically no relation to such discussions
in cognitive science and philosophy. Therefore one may, in principle, ignore usage of this misleading terminology and just say that in the Bell
type experiments randomness of a source of entangled systems is combined with random selection of experimental setting, e.g., in the form
of angles, orientations of PBSs. The latter is realized with the aid of (pseudo)-random generators. Of course, experimenter has the freedom of choice
of these random generators (as well as she has also the freedom to do or not the experiment at all). However, after generators were selected 
randomness involved in the Bell experiment (in fact, a few experiments)  became of purely physical nature, i.e., the mental element is totally excluded. Nevertheless, one may argue that 
an experimenter can in principle interrupt the production of random numbers by changing at the arbitrary instant of time the parameters of
random generators and here her free will will again play a role. In principle, we can accept this position and say that our model includes 
the mental randomness produced by brain's functioning. However, if we consider the brain as a physical system which activity is based 
on spikes produced by neurons (or/and electromagnetic fields generated in the brain), then we again obtain the pure physical account 
of randomness in the Bell type experiments. Free will would play an exceptional role only if one accepts that human consciousness is not 
reduced to physical processes in the brain. Such a position is still valuable, e.g., \cite{Padic}. By accepting it we would agree 
that Kolmogorovization is successful only as the result of accounting of combined physical-mental randomness.      

G. `t Hooft rejects the existence of free will. We mention that his position is close to the position of the majority of cognitive scientists and philosophers
working on the problem of free will \cite{Free}. The common opinion is that free will is one of the traces of the God-based picture of the world. The brain 
is a sort of deterministic device (which, of course, contains various noisy signals) and all acts of free will are dynamically predetermined 
by brain's state in previous instances of time. We state again that our model need not be based on such a position. We are fine by assuming 
the existence of a nonphysical pure mental elements or even God, in any event the classical probability model for Bell's experiment is well defined.

\bigskip 

{\bf Acknowledgment:}
This work was partially supported by MPNS COST Action MP1006 (Fundamental Problems in Quantum Physics), by visiting fellowship 
 to  Institute for Quantum Optics and Quantum Information, Austrian Academy of Sciences, and by the grant of Linnaeus University,
Mathematical Modeling of Complex Hierarchic Systems.

\end{document}